\documentclass[12pt]{article}
\usepackage{latexsym}

\newcommand{\xxx}[1]{ [#1]}
\newcommand{\mysection}[1]{\section{#1}
   \hspace{0.8cm}\setcounter{equation}{0}}
\renewcommand{\theequation}{\arabic{section}.\arabic{equation}}
\newcommand{\myappendix}{\appendix
   \renewcommand{\theequation}{\Alph{section}.\arabic{equation}}
   \vspace{30pt} \noindent {\Large \bf Appendix}}
\def\thefootnote{\fnsymbol{footnote}}
\def\notd{\not{\hspace{-.03in}\pp}}
\def\notp{\not{\hspace{-.03in}p}}

\def\notD{\not{\hspace{-.05in}D}}
\def\notA{\not{\hspace{-.05in}A}}
\newcommand{\vev}{{\it vev}}
\def\tG{{\tilde G}}

\def\tF{{\tilde F}}

\def\bl{\bar{\lambda}}

\def\r{\right|}
\def\l{\left.}
\def\[{\left [}
\def\]{\right ]}
\def\({\left (}
\def\){\right )}
\def\|{\left |}
\def\lbr{\left\{}
\def\rbr{\right\}}
\def\pp{\partial}

\def\R{{\bar{R}}}

\def\ux{$U(1)_X$}

\def\I{\bar{I}}
\def\t{\bar{t}}
\def\T{\bar{T}}

\def\m{\bar{m}}
\def\M{\bar{M}}

\def\Tr{{\rm Tr}}

\def\im{{\rm Im}}
\def\re{{\rm Re}}
\def\eV{{\rm eV}}
\def\MeV{{\rm MeV}}
\def\GeV{{\rm GeV}}

\newcommand{\beq}{\begin{equation}}
\newcommand{\eeq}{\end{equation}}
\newcommand{\bea}{\begin{eqnarray}}
\newcommand{\eea}{\end{eqnarray}}
\def\bF{\bar{F}}
\def\G{{\cal G}}
\def\f{\bar{f}}

\def\L{{\cal L}}
\newcommand{\superint}{\int \diff^{4}\theta \, }

\newcommand{\lowest}{|_{\theta =\bar{\theta}=0}}
\newcommand{\diff}{\mbox{d}}

\def\D{{\cal D}}

\def\notcD{\not{\hspace{-.05in}\D}}

\newcommand{\WaWa}{{\cal{W}}^{\alpha}{\cal{W}}_{\alpha}}

\newcommand{\DaDa}{{\cal D}^2}
\newcommand{\DbDb}{{\bar{\cal D}}^2}

\newcommand{\Wa}{{\cal W}^{\alpha}}
\newcommand{\Wbb}{{\overline{\cal W}}_{\dot{\beta}}}
\newcommand{\Wc}{{\cal W}_{\alpha}}

\newcommand{\Wd}{{\overline{\cal W}}^{\dot{\beta}}}

\def\pp{\partial}

\def\Del{\Delta}

\def\del{\delta}

\def\u{\bar{u}}

\newcommand{\myref}[1]{(\ref{#1})}
\newcommand{\beqa}{\begin{eqnarray}}
\newcommand{\eeqa}{\end{eqnarray}}
\newcommand{\nnn}{ \nonumber \\ }

\newcommand{\hc}{ + {\rm h.c.}}
\newcommand{\mhc}{ - {\rm h.c.}}

\newcommand{\uone}{$U(1)$}

\newcommand{\myvev}[1]{{\langle #1 \rangle}}
\newcommand{\bigvev}[1]{{\left\langle #1 \right\rangle}}

\newcommand{\half}{{1 \over 2}}

\newcommand{\ddd}{\nnn && \quad}
\newcommand{\mmm}{\nnn & &}

\def\hG{\hat G}
\def\hQ{\hat Q}
\def\hq{\hat q}

\def\eee{\nonumber \\ &=&}

\def\tq{{\tilde q}}
\def\tQ{{\tilde Q}}

\newcommand{\mysec}[1]{Section \ref{#1}}
\newcommand{\myapp}[1]{Appendix \ref{#1}}
\def\bpi{\mathbf{\Pi}}
\def\bfpi{\mbox{\boldmath$\pi$}}

\def\bfs{\mbox{\boldmath$\sigma$}}
\def\bfm{\mathbf{M}}

\def\bff{\mathbf{F}}
\def\bfq{\mathbf{q}}
\def\bfa{\mathbf{a}}
\def\bfk{\mathbf{K}}
\def\bfbf{\mathbf{\bF}}
\def\bfone{\mathbf{1}}
\def\bfP{\mathbf{P}}\def\bfS{\mathbf{S}}
\def\mG{m_{3\over2}}

\begin{document}

\begin{titlepage}

\hfill   LBNL-59025

\hfill   UCB-PTH-05/36


\hfill   October 2005

\begin{center}

\vspace{18pt}
{\bf Is the Universal String Axion the QCD Axion?}\footnote{This
work was supported in part by the
Director, Office of Science, Office of High Energy and Nuclear
Physics, Division of High Energy Physics of the U.S. Department of
Energy under Contract DE-AC02-05CH11231, in part by the National
Science Foundation under grant PHY-0098840.}

\vspace{18pt}

Mary K. Gaillard\footnote{E-Mail: {\tt MKGaillard@lbl.gov}}
{\em and}
Ben Kain\footnote{E-Mail: {\tt bkain@socrates.berkeley.edu}}
\vskip .01in
{\em Department of Physics, University of California 
and \\ Theoretical Physics Group, Bldg. 50A5104,
Lawrence Berkeley National Laboratory \\ Berkeley,
CA 94720 USA}
\vskip .03in

\vspace{18pt}

\end{center}

\begin{abstract} 

We consider the class of effective supergravity theories from the
weakly coupled heterotic string in which local supersymmetry is broken
by gaugino condensation in a hidden sector, with dilaton stabilization
achieved through corrections to the classical dilaton K\"ahler
potential. If there is a single hidden condensing (simple) gauge
group, the axion is massless (up to contributions from higher
dimension operators) above the QCD condensation scale. We show how the
standard relation between the axion mass and its Planck scale coupling
constant is modified in this class of models due to a contribution to
the axion-gluon coupling that appears below the scale of supersymmetry
breaking when gluinos are integrated out.  In particular there is a
point of enhanced symmetry in parameter space where the axion mass is
suppressed.  We revisit the question of the universal axion as the
Peccei-Quinn axion in the light of these results, and find that the
strong CP problem is avoided in most compactifications of the weakly
coupled heterotic string.

\end{abstract}

\end{titlepage}

\newpage
\renewcommand{\thepage}{\roman{page}}
\setcounter{page}{2}
\mbox{ }

\vskip 1in

\begin{center}
{\bf Disclaimer}
\end{center}

\vskip .2in

\begin{scriptsize}
\begin{quotation}
This document was prepared as an account of work sponsored by the United
States Government. Neither the United States Government nor any agency
thereof, nor The Regents of the University of California, nor any of their
employees, makes any warranty, express or implied, or assumes any legal
liability or responsibility for the accuracy, completeness, or usefulness
of any information, apparatus, product, or process disclosed, or represents
that its use would not infringe privately owned rights. Reference herein
to any specific commercial products process, or service by its trade name,
trademark, manufacturer, or otherwise, does not necessarily constitute or
imply its endorsement, recommendation, or favoring by the United States
Government or any agency thereof, or The Regents of the University of
California. The views and opinions of authors expressed herein do not
necessarily state or reflect those of the United States Government or any
agency thereof of The Regents of the University of California and shall
not be used for advertising or product endorsement purposes.
\end{quotation}
\end{scriptsize}

\vskip 2in

\begin{center}
\begin{small}
{\it Lawrence Berkeley Laboratory is an equal opportunity employer.}
\end{small}
\end{center}

\newpage
\renewcommand{\thepage}{\arabic{page}}
\setcounter{page}{1}

\def\thefootnote{\arabic{footnote}} \setcounter{footnote}{0}
\mysection{Introduction} 
As observed by Banks and Dine~\cite{bd1}, in a supersymmetric Yang
Mills theory with a dilaton chiral superfield that couples universally
to Yang-Mills fields there is a residual R-symmetry in the effective
theory for the condensates of a strongly coupled gauge sector,
provided that there is a single condensation scale governed by a
single $\beta$-function, there is no explicit R-symmetry breaking by
fermion mass terms in the strongly coupled sector, and the dilaton $S$
has no potential. The latter requirement is met in effective
supergravity from string theory, and explicit realizations of this
scenario have been constructed~\cite{bgw,ggm} in the context of the
weakly coupled heterotic string.  Since the axion is massless above
the QCD condensation scale in these scenarios, it is a natural
candidate for the Peccei-Quinn axion.

In~\cite{bgw,ggm} gaugino and matter condensation was studied for
gauge sectors that have no dimension-two gauge invariant operators. In
these effective theories, two-condensate models have a point of
enhanced symmetry where the condensing gauge sectors $\G_a$ have the
same beta-function coefficients $b_a$, and the axion mass is
proportional to $|b_1 - b_2|$.  Here we are interested in the case
where one condensing gauge group $\G_Q$ is $SU(N_c)$ with a $U(N)$
flavor symmetry for quark supermultiplets.  In the following section
we construct supersymmetric models for this case. This has the
advantages that all the symmetries are manifest and the effective
Lagrangian is highly constrained by supersymmetry.  In \mysec{chiral}
we extend the methods of~\cite{bgw} to the case of $\G_Q$ and show
that in the rigid supersymmetry limit $m_P\to\infty$ we recover known
results~\cite{dds}. In \mysec{toy} we include a condensing hidden
sector with a gauge group $\G_c$ of the class studied in~\cite{bgw}
and show that the point of enhanced symmetry in this case corresponds
to
\beq b_c = {N_c\over8\pi^2}.\label{sympt}\eeq
Although this is not a realistic model for QCD, where the
condensation scale is far below the scale of supersymmetry breaking,
we recover the effective potential for light pseudoscalars by
identifying the $q\bar q$ pseudoscalar bound states with the
F-components of the quark condensate superfields, and show that the
presence of the symmetric point \myref{sympt} is reflected in the 
axion mass.

In \mysec{qcd} we consider a more realistic model in which the QCD
gauge and matter degrees of freedom are unconfined at the
supersymmetry-breaking scale. We show that R-symmetry is not broken by
the large gaugino (and squark) masses because the masslessness of the
axion leaves the phase of the gaugino masses undetermined. We argue
that this implies that the symmetries of the effective theory at the
supersymmetry-breaking scale must be reflected in the effective
quark-gluon theory just above the QCD condensation scale, implying a
correction to the axion-gluon coupling.  This result is confirmed by
an explicit calculation of the heavy gaugino loop contribution in
\myapp{anomap}, and we recover the result of \mysec{susy} for the
axion mass.

For QCD with $N_c = 3$, the point of enhanced symmetry \myref{sympt} has
\beq b_c = {3\over8\pi^2} = .038,\eeq
which is in the preferred range $.3\le b_c\le .4$ found in studies of
electroweak symmetry-breaking~\cite{gn2} and of dark matter
candidates~\cite{bhn} in the context of the models considered here.
As a consequence the axion mass is suppressed and higher dimension
operators~\cite{bd1,bg} might lead to strong CP violation.
We address this question in \mysec{cpv}.

Our results are summarized in \mysec{conc}.  Here we use the linear
supermultiplet formulation for the dilaton superfield, but we expect
that our results can be reproduced in the chiral multiplet formulation.

\mysection{Supersymmetric  models}\label{susy}
The supersymmetry breaking models of~\cite{bgw,ggm} are based on
strongly coupled hidden sector gauge groups of the form
$\prod_a\G_a$. The generalization to supergravity~\cite{tom,bg89} of
the VYT effective action~\cite{vyt} is obtained by introducing 
composite field operators $U_a$ and $\Pi^\alpha_a$ that are $\G_a$-charged
gauge and matter condensate chiral
superfields, respectively:
\beq U_a\simeq {\Wa_a\Wc^a},\qquad \Pi_a^\alpha\simeq
\prod_A\(\Phi_a^A\)^{n^A_{\alpha,a}},\label{condef} \eeq 
and by matching the anomalies of the effective theory to those of the
underlying theory.  The Lagrangian\footnote{We work in the K\"ahler
\uone\, superspace formalism of~\cite{bgg}.}
\bea \L_{VYT} = {1\over8}\superint\,{E\over R}\sum_a
U_a\[b'_a\ln(e^{-K/2}U_a) + \sum_\alpha b^\alpha_a\ln\Pi^\alpha\] +
{\rm h.c.},\label{vyt} \eea
has the correct anomaly structure under K\"ahler \uone\, R-symmetry:
\bea \lambda_\alpha^a &=& \l\Wc^a\r\to
e^{i\alpha/2}\lambda_\alpha^a,\qquad \chi_\alpha^A =
{1\over\sqrt{2}}\l\D_\alpha\Phi^A\r\to e^{-i\alpha/2}\chi_\alpha^A,
\nnn U_a&\to& e^{i\alpha}U_a,\qquad \Pi^\alpha_a
\to\Pi^\alpha_a,\label{kuone}\eea
conformal transformations:
\bea \lambda_a&\to&
e^{3\sigma/2}\lambda_a,\qquad \Phi^A\to e^{\sigma}\Phi^A,\qquad etc.,
\nnn U_a&\to& e^{3\sigma}U_a,\qquad \Pi^\alpha_a
\to e^{d_a^\alpha\sigma}\Pi^\alpha_a,\qquad d_a^\alpha
= \sum_A n^A_{\alpha,a},\label{conft}\eea
and modular (T-duality) transformations that involve functions
of the moduli chiral supermultiplets $T^I$:
\bea\lambda_a&\to& e^{-i\im F/2}\lambda_a,\qquad \chi^A\to e^{-F^A
+i\im F/2}\chi^A, \qquad F^A = \sum_I q^A_I F^I(T^I),
\qquad F = \sum_I F^I,\nnn U_a&\to& e^{-i\im
F}U_a,\qquad \Pi^\alpha_a \to e^{-F^\alpha_a}\Pi^\alpha_a, \qquad
F^\alpha_a = \sum_A n^A_{\alpha,a}F^A,\label{tmod}\eea
provided the conditions 
\beq b'_a = {1\over8\pi^2}\(C_a - \sum_AC_a^A\) ,\;\;\;\; 
\qquad b^\alpha_a = \sum_{A\in\alpha}{C^A_a\over4\pi^2d_a^\alpha},
\label{cond}\eeq
are satisfied, where $C_a(C_a^A)$ is the quadratic Casimir in the
adjoint ($\Phi^A$) representation of $\G_a$.  In the heterotic
superstring theory, the anomaly under \myref{tmod} is canceled by
moduli dependent string loop threshold corrections~\cite{thresh} and a
four-dimensional version~\cite{gsterm} of the
Green-Schwarz~\cite{GS84} term.  In effective supergravity theories
with an anomalous gauge group \ux, there are additional anomaly
matching conditions~\cite{ggm} and an additional Green-Schwarz
term~\cite{UXR} to restore the \ux\, invariance of the underlying
string theory. The composite chiral superfields $\Pi^\alpha_a$ are
invariant under the nonanomalous symmetries, and may be used to
construct an invariant superpotential~\cite{bgw,ggm}. Provided there
are no invariant chiral fields of dimension two, and no additional
global symmetries (such as chiral flavor symmetries), the dynamical
degrees of freedom associated with the composite fields \myref{condef}
acquire masses~\cite{yy} larger than the condensation scale
$\Lambda_a$, and may be integrated out, resulting in an effective
theory constructed as described above with the composite fields taken
to be nonpropagating; that is, they do not appear in the K\"ahler
potential. The dynamical axion is massless if there is a single
condensate, or if the condensing gauge groups all have the same
$\beta$-function coefficient $b_a$, defined by
\beq {\pp g_a(\mu)\over\pp\ln\mu} = - {3b_a\over2}g^3_a(\mu).\label{ba}\eeq
In this case there is a nonanomalous R-symmetry: the axion shift
compensates the anomaly arising from the transformation
\myref{kuone}. The symmetry is broken if there are condensing gauge
groups with different coefficients $b_a$.  For example in the case of
two condensates with coefficients $b_1>b_2$ and condensation scales
$\Lambda_1\gg\Lambda_2$, the axion mass is approximately given
by~\cite{bgw}
\beq m_a\approx {3\myvev{\ell}\sqrt{b_1}\over b_2}\[(b_1 - b_2){\Lambda_2
\over\Lambda_1}\]^{3\over2}\mG,\label{b1b2}\eeq
where $\ell$ is the dilaton field in the linear multiplet formulation.
In the classical approximation $\myvev{\ell} = g^2_s/2\approx.25$,
with $g_s$ the string coupling constant. Generally we expect
$\myvev{\ell}\sim 1$ when string nonperturbative~\cite{shenk} and/or
field theoretic quantum corrections~\cite{bd1,Derendinger95,gn} to the
dilaton K\"ahler potential are invoked to assure dilaton
stabilization~\cite{casas}.

In the case that there is just one hidden sector condensing gauge
group $\G_c$, the axion remains massless above the scale
$\Lambda_{QCD}$ of quark and gluon condensation in the standard model,
and is a candidate for the Peccei-Quinn axion.  However in this case
there is an enlarged symmetry in the limit of one or more massless
quarks, leading to light condensates $m_\pi<\Lambda_{QCD}$, and the
result \myref{b1b2} is modified.

In order to more closely model QCD, in this section we consider
supergravity models with a strongly coupled gauge group $\G_c\otimes
SU(N_c)$ with $N\equiv N_f < N_c$ vector-like flavors. We first
extend the construction of the effective VYT action to this case, and
show that in the flat SUSY limit it reduces to the results~\cite{dds} based
on the holomorphy of the superpotential. 

\subsection{The VYT action for $SU(N_c)$ with 
chiral flavor symmetry}\label{chiral}
We have N ``quark'' and N ``anti-quark'' chiral supermultiplets $Q^A$
and $Q^c_A$, respectively.  We take the quark condensates to be the
matrix-valued ``meson'' superfield $\bpi^A_B = Q^A Q^c_B$. We do not
assume {\it a priori} that these are static fields.  For $\G_a =
SU(N_c)\equiv\G_Q$ we take
\beq \L^Q_{VYT} = {1\over8}\superint{E\over R}U_Q\[b'_Q\ln(e^{-K/2}U_Q) +
b^\alpha_Q\ln(\det\bpi)\]\hc\label{lvyt}\eeq
For the elementary fields we have $C_Q = N_c$, $C^A_Q = \half,$
$\sum_A C^A_Q = N$.  Under K\"ahler \uone\, R-symmetry \myref{kuone}
anomaly matching requires
\beq b'_Q = {1\over8\pi^2}\(N_c - N\),\label{bpq}\eeq
and under the conformal transformation \myref{conft} with $\bpi \to
e^{2\sigma}\bpi$, we require
\beq 3b'_Q + 2N b^\alpha_Q = {1\over8\pi^2}\(3N_c - N\) = 3
b_Q,\label{conf}\eeq
where $b_Q$ is the $\beta$-function coefficient as defined in
\myref{ba}.  Putting these together gives \myref{bpq} and
\beq b^\alpha_Q = {1\over8\pi^2}, \qquad b_Q = b'_Q +
{2N\over3}b^\alpha_Q,\label{baq}\eeq
in agreement with the general result \myref{cond} with $d_Q^\alpha = 2N$
for $\pi_Q^\alpha = \det\bpi.$
If $Q,Q^c$ have modular weights $\bfq_I,\bfq^c_I$, under T-duality
\myref{tmod}
\beq Q\to e^{-\bff^Q} Q,\quad Q^c\to e^{-\bff^{Q^c}}Q^c,\qquad \bff^Q
= \sum_I\bfq_I F^I(T^I), \quad \bfq_I = {\rm diag}(q_I^1,\ldots
q_I^N),\eeq
and similarly for $\bff^{Q^c}$. The modular anomaly matching condition
\beq  b'_Q + b^\alpha_Qq^\alpha_I =
{1\over8\pi^2} \lbr C_Q + \sum_{A}C^A_Q\[2q^A_I + 2(q^c)^A_I-
1\]\rbr,\qquad q^\alpha_I = \sum_A\(q^A_I + (q^c)^A_I\), \eeq
is also satisfied by \myref{bpq} and \myref{baq}.  Finally, like the
underlying theory, \myref{lvyt} is invariant under flavor
$SU(N)_L\otimes SU(N)_R$, while under chiral $U(1)$ transformations
\beq Q\to e^{i\beta}Q,\quad Q^c\to e^{i\beta}Q^c\label{uone}\eeq
the anomaly matching condition
\beq 2N b^\alpha_Q = \sum_A{C^A_Q\over4\pi^2}\eeq
is also satisfied.

For the superpotential we take 
\beq W(\bpi) = \eta^{-2}\Tr\[\prod_I\eta_I^{2\bfq_I}
\bpi\prod_J\eta_J^{2\bfq_J^c}\bfm\], \eeq
where $\bfm$ is the mass matrix and $\eta_I = \eta(T^I)$, $\eta =
\prod_I\eta_I.$ The component Lagrangian for the effective theory with
just this $SU(N_c)$ condensate can easily be inferred from the results
of~\cite{bgw}.  Solving the equation of motion for $\re F^Q$ gives
\bea u_Q &=& e^{i\omega_Q}\lambda_Q[\det(\bpi\bpi^{\dag})]^{-1/2(N_c -
N)}, \nnn \lambda_Q &=& e^{-1}e^{[k + K(\bpi)]/2}\prod_I\[2\re
t^I|\eta(t^I)|^4\]^{(b-b'_Q)/2b'_Q}|\eta(t^I)|^{-
2b^\alpha_Qq^\alpha_I/b'_Q} \Lambda_Q^{(3N_c-N)/(N_c -
N)},\label{uq}\eea
where 
\beq \Lambda_a = \exp(- s(\ell)/3b_a),\qquad s(\myvev{\ell}) =
g^{-2}_s,\label{lambda}\eeq
is defined as the scale at which the one loop running coupling $g_a(\mu)$
blows up:
\beq g_a^{-2}(\Lambda_a) = g_s^{-2} + 3b_a\ln(\Lambda_a/m_P),\eeq
in reduced Planck mass units, $m_P = 1$, that we use
throughout.\footnote{A correction that accounts for the fact that the
string scale differs from the Planck scale by $\mu_s = g_s m_P$ is
encoded in the $e^k$ factor in $|\u_a|^2$.}  To compare with previous
results~\cite{dds} we take the rigid SUSY limit, and neglect the
moduli and the dilaton; $s(\ell)\to g_0^{-2}$.  Then the
superpotential reduces to the standard VY one:
\beq W(U_Q) = {1\over4}U_Q\[g_0^{-2} + b'_Q\ln(U_Q) +
b^\alpha_Q\ln(\det\bpi)\].\label{wvy}\eeq
Keeping $U_Q$ static and imposing the equation of motion for the
auxiliary field $F^Q$ gives the potential
\beq - V = \Tr\[\bfbf^\pi\bfk''\bff^\pi + \lbr\bff^\pi\(\bfm +
{1\over4}b^\alpha_Q u_Q\bpi^{-1}\) + \rm h.c.\rbr\], \qquad u_Q =
e^{-1}\({\Lambda_Q^{3N_c-N}\over\det\bpi}\)^{1/(N_c -
N)}\label{susyv}\eeq
where $\bfk''$ is the (tensor-valued) K\"ahler metric for
$\bpi$. Since $\pp(\det\bpi)^p/\pp\bpi = p\bpi^{-1}\det\bpi^p,$
the potential \myref{susyv} is derivable from the following
superpotential for the dynamical superfield $\bpi$:
\beq W_{\bpi} = \Tr(\bfm\bpi) - {(N_c -   N)\over32\pi^2e}
\({\Lambda_Q^{3N_c-N}\over(\det\bpi)}\)^{1/(N_c - N)},\label{wnp}\eeq 
which, up to a factor\footnote{The factor $e$ comes from the fact that
we take the derivative of $\int U\ln U$, while the authors of
\cite{dds} start with $\int <\lambda\lambda>\ln\Lambda$ and determine
$<\lambda\lambda>$ from threshold matching. The minus sign comes from
the convention of~\cite{bgg}: $u\sim\l\Wa\Wc\r = - \lambda\lambda$.}
$-2/e$, is the superpotential found in \cite{dds}. 

We may also consider the case -- more closely resembling QCD -- where
only $n<N$ chiral supermultiplets have masses below the condensation
scale $u_Q^{1/3}\sim\Lambda_Q$, while $m = N-n$ chiral supermultiplets
have masses $M_A$ above that scale. The latter decouple at scales below
their masses, which explicitly break the nonanomalous $U(N)_L\otimes
U(N)_R$ symmetry to a $U(n)_L\otimes U(n)_R$ symmetry if $m = N - n$
quarks are massive. They do not contribute to the chiral anomaly at 
the $SU(N_c)$ condensation scale.  To account for these effects we
replace \myref{lvyt} by
\beq \L^Q_{VYT} = {1\over8}\superint{E\over R}U_Q\lbr b'_n\ln(e^{-K/2}U_Q)
+ b^\alpha_Q\[\ln(\det\bpi_n) - \sum_{A=1}^m\ln
M_A\]\rbr\hc,\label{lvyt3}\eeq
where $b'_n = (N_c - n)/8\pi^2$ and $\bpi_n$ is an $n\times n$
matrix-valued composite operator constructed only from light quarks.
\myref{lvyt3} can be formally obtained from \myref{lvyt} by
integrating out the heavy quark condensates as follows. As the
threshold $M_A$ is crossed, set $\det\bpi_{n+A}\to \pi^A\det\bpi_{n +
A -1}$ and take the condensate $\pi^A \sim Q^A Q_A^c$ to be static:
$K(\bpi_{n+A})\to K(\bpi_{n + A -1})$. Then including the
superpotential term $W(\pi^A) = - M_A\pi^A$, the equation of motion
for $F^A$ gives $\pi^A = e^{-K/2}u_Q/32\pi^2M_A$, giving \myref{lvyt3}
up to some constant threshold corrections.  The flat SUSY analogue of
\myref{wvy} is now
\beq W = {1\over4}U_Q\lbr g_0^{-2} + b'_n\ln(U_Q)
+ b^\alpha_Q\[\ln(\det\bpi) - \sum_{A=1}^m\ln
M_A\]\rbr,\label{wvy2}\eeq
and we recover \myref{susyv}--\myref{wnp} with now
\beq \Lambda_Q = e^{-1/3b_n g^{2}}\prod_{A=1}^m M_A^{b_3/3b_n}, \qquad
3b_n = {3N_c - n\over8\pi^2} = 3b'_n + 2n b_3,\eeq
which corresponds to running $g^{-2}(\mu)$ from $g^{-2}(1)= g_0^{-2}$
to $g^{-2}(\Lambda_Q)= 0$ using the $\beta$-function coefficient
$(3N_c - n - A)/8\pi^2$ for $m_{A}\le \mu\le m_{A+1}$, again in
agreement with the results of nonperturbative flat SUSY
analyses~\cite{dds}.

\subsection{Supergravity with strongly coupled 
$\G_c\otimes SU(N_c)$}\label{toy}
We consider the supergravity action defined by
\beq \L = \L_K + \L_P + \L_{VYT} + \L_{GS} + \L_{Th},\eeq
where $\L_{GS}$ and $\L_{Th}$ contain, respectively, the Green-Schwarz
terms and threshold corrections discussed above, 
\beq \L_P = \half\superint{E\over R}e^{K/2}W, \eeq
is the superpotential term, and
\beq \L_K = \superint E\[- 3 + 2L s(L)\]\eeq
contains the locally supersymmetric extension of the Einstein and
Yang-Mills terms as well as the kinetic terms for matter through the
K\"ahler potential
\beq K = k(L) + K(\bpi) - \sum_I\ln\(T^I + \T^{\I}\), \qquad  k'(L) 
+ 2L s'(L) = 0.\eeq
The Yang-Mills term arises from the modified linearity conditions~\cite{bgg}
defined by the chiral projections of the real field $L$:
\beq \(\DbDb - 8R\)L = - \sum_a\(\WaWa\)_a, \qquad \(\DaDa - 8\R\)L
= -\sum_a\(\Wd\Wbb\)_a.\label{lin}\eeq
Below the condensation scale $\Lambda_a$ we make the replacement
$\(\WaWa\)_a\to U_a$ in \myref{lin}.  The VYT term is given by
\myref{lvyt}--\myref{baq} for $a=Q$.  For $\G_c$ we follow \cite{bgw}
and take dimension-three operators for the $\Pi^\alpha_c$ in
\myref{vyt}, giving the anomaly matching conditions:
\bea b_c &=& b'_c + \sum_\alpha b^\alpha_c = {1\over8\pi^2}\(C_c -
\textstyle{1\over3}\sum_A C^A_c\), \nnn b'_c &=& {1\over8\pi^2}\(C_c -
\sum_A C^A_c\), \qquad  b^\alpha_c = {1\over12\pi^2}\sum_A C^A_c.
\label{matchc}\eea
For the superpotential we now take 
\beq W = \sum_\alpha c_\alpha\prod_I\eta_I^{2(q^\alpha_I -
1)}\Pi^\alpha_c + \eta^{-2}\Tr\[\prod_I\eta_I^{2\bfq_I}
\bpi\prod_J\eta_J^{2\bfq_J^c}\bfm\], \eeq
and we approximate the K\"ahler potential for $\bpi$ by\footnote{We could
take a more general form of the K\"ahler potential, {\it e.g.,} $K\sim
\Tr\bpi\bpi^{\dag}/\Tr\bpi$, $\Tr\bpi\bpi^{\dag}/(\det{\bpi})^{1/N}$
or $(\Tr\bpi\bpi^{\dag})^{\half}$; these would give essentially the
same result with $\mu^2\sim v$.}
\beq K(\bpi) = \mu^{-2}\Tr\[\prod_I(T^I +
\T^{\bar I})^{-\bfq_I}\bpi\bpi^{\dag} \prod_J(T^J +
\T^{\bar J})^{-\bfq^c_J}\].\eeq
We can make a holomorphic field redefinition such that\footnote{We
took the superpotential to be modular invariant; in the context of
string theory, this implies an implicit assumption that the \vev\,
that induces the mass term does not break modular invariance. 
This need not be the case for the more realistic QCD model discussed 
in \mysec{qcd} where the masses are generated by the MSSM Higgs \vev\,
$v_H\ll\Lambda_c$; as a result there can be small corrections to the
effective theory of order $v_H/\Lambda_c$. These do not affect our 
conclusions which depend only on the residual R-symmetry above
the QCD confinement scale and, for CP violation discussed in \mysec{cpv},
on unbroken T-duality above $\Lambda_c$.}  it
becomes obvious that the moduli are still stabilized at self dual
points with vanishing F-terms, namely
\beq \bpi = \prod_I \eta_I^{-2q_I}\bpi'\prod_J \eta_J^{-2q^c_J},
\label{newpi2}\eeq
and then drop the prime. Then the last line in \myref{uq} becomes
\beq \lambda_Q = e^{-1}e^{[k + K(\bpi)]/2}\prod_I\[2\re
t^I|\eta(t^I)|^4\]^{(b - b'_Q - b^\alpha_Qq^\alpha_I)/2b'_Q}
\Lambda_Q^{(3N_c-N)/(N - N_c)},\label{lq}\eeq
and the K\"ahler potential and superpotential are now
\bea K(\bpi) &=&  \mu^{-2}\Tr\lbr\prod_I\[(T^I +
\T^{\bar I})|\eta_I|^4\]^{-\bfq_I}\bpi\bpi^{\dag} \prod_J\[(T^J +
\T^{\bar J})|\eta_J|^4\]^{-\bfq^c_J}\rbr,\nnn W &=& \sum_\alpha
c_\alpha\prod_I\eta_I^{2(q^\alpha_I - 1)}\Pi^\alpha_c +
\eta^{-2}\Tr\(\bpi\bfm\).\label{pikw} \eea
Then the moduli derivatives $K_I(\bpi)$ and $\pp_I[e^K W(\bpi)]$ with
the new $\bpi$ variables fixed vanish at the self dual points.  To
study the potential for the other fields we may set the moduli at
their ground state values. We could neglect the moduli altogether in
this toy model since modular invariance and K\"ahler R-symmetry give
the same anomaly matching conditions.  However it is also interesting
to check that mixing of the axion with $\im t^I = \im\l T^I\r$ makes
no difference. Since we are only concerned with the phase $N\del$ of
$\det\bfm$, we make the simplifying assumption that $\bfm = m
e^{i\del}\bfone$. Then $\myvev{\bpi} = v e^{i\phi_\pi}\bfone$, and 
a convenient parameterization is 
\bea \bpi &=& \bfS e^{i\bfP}, \qquad \bfS = v\[1 + {\mu\over
v\sqrt{2}}\(\bfs + {\sigma_0\over\sqrt{N}}\)\],\qquad \bfP = \phi_\pi
+ {\mu\over v\sqrt{2}}\(\bfa + {a_0\over\sqrt{N}}\),\nnn \det\bpi &=&
\det\bfS e^{i(N\phi_\pi + \mu\sqrt{N}a_0/v\sqrt{2})}, \qquad
\Tr|\bpi|^2 = \Tr\bfS^2, \qquad \Tr|\bpi|^{-2} = \Tr\bfS^{-2},\nnn
\myvev{\bfs} &=& \myvev{\sigma_0} = \myvev{\bfa} = \myvev{a_0} = 0,
\qquad \bfs = \sqrt{2}\sum_i T_i\sigma^i,\qquad \bfa = \sqrt{2}\sum_i
T_i a^i,
\label{newpi4}\eea
where $T_i$ is a generator of flavor $SU(N)$ in the fundamental
representation:
\beq \Tr T_i = 0,\qquad \Tr T_i^2 = \half.\label{norm}\eeq  
The component Lagrangian for bosons then takes the form
\beq \L = \L_{KE} - V,\eeq
The potential is
\bea V &=& {k'\over16\ell}\left|\rho_c e^{i\omega_c}(1 + b_c\ell) +
\lambda_Q|\det\bpi|^{-1/(N_c-N)}e^{i\omega_Q}(1 + \ell b'_Q) -
4e^{K/2}\ell\eta^{-2}m e^{i\del}\Tr\bpi\right|^2 \mmm -
{3\over16}\left|b_c\rho_c e^{i\omega_c} + b'_Q\lambda_Q
|\det\bpi|^{-1/(N_c-N)}e^{i\omega_Q} - 4e^{K/2}\eta^{-2}m
e^{i\del}\Tr\bpi\right|^2\mmm + \sum_I K_{I\bar I}F^I\bF^{\bar I} +
\Tr\bfbf^\pi\bfk''\bff^\pi, \nnn K_{I\bar I}\bF^{\bar I} &=& {4\re
t^I\zeta(t^I) + 1\over8\re t^I}
\[(b - b_c)\rho_c e^{i\omega_c} + (b - b'_Q)
\lambda_Q|\det{\bpi}|^{-1/(N_c-N)}e^{i\omega_Q}\right.\ddd\qquad
\qquad\l + 4e^{K/2}\eta^{-2}
m e^{i\del}\Tr\bpi\] \nnn (\bfbf^\pi\bfk'')^A_B &=&
{1\over4}\lbr{\mu^{-2}}\overline{\bpi}^A_B\[b_c\rho_c e^{i\omega_c} +
b'_Q\lambda_Q|\det{\bpi}|^{-1/(N_c-N)}e^{i\omega_Q} -
4e^{K/2}\eta^{-2}m e^{i\del}\Tr\bpi\] \right.\ddd\qquad\qquad\l
- b^\alpha_Q\lambda_Q\(\bpi^{-1}\)^A_B
|\det{\bpi}|^{-1/(N_c-N)}e^{i\omega_Q} - 4e^{K/2}\eta^{-2} m\del^A_B
e^{i\del}\rbr,\eea
which can be written in the form
\bea V &=& A_1(\bfS) + A_2(\bfS)\cos(\omega'_c - \omega'_Q) +
m_0\Tr\lbr e^{-i\mu\bfa/v\sqrt{2}}\[A_3(\bfS)e^{i\omega'_c} +
A_4(\bfS)e^{i\omega'_Q}\] \hc\rbr\mmm + m_0^2A_5(\bfS,\bfa^2),
\qquad \omega'_a = \omega_a - \del - \phi_\pi - \nu a_0 -
i\ln(\eta/\bar\eta),\qquad \nu = {\mu\over v\sqrt{2N}},\label{va}\eea
where the $A_i$ are real and independent of the condensate phases, and
\beq m_0 = e^{K/2}|\eta|^{-2}m\eeq
is the quark mass.  Using the parameterization \myref{newpi4}, the
kinetic energy term is, dropping a total derivative,
\bea \L_{KE} &=& {k'\over4\ell}B^m B_m - b_c\tilde\omega_c\nabla^m
B^c_m - b'_Q\tilde\omega_Q\nabla^m B^Q_m + i{b\over2}B^m\sum_I\(\ell'^I\pp_m
t^I\mhc\)\mmm - \half\(\pp_m\bfs\pp^m\bfs + \pp_m\sigma_0
\pp^m\sigma_0 + \pp_m\bfa\pp^m\bfa + \pp_m a_0\pp^m a_0\)
 - K_{I\bar J}\pp t^I\pp\t^{\bar J},\nnn
\tilde\omega_c &=& \omega_c - i\ln(\eta/\bar\eta),\quad \tilde\omega_Q
= \omega_Q - i\ln(\eta/\bar\eta) + {N\over N_c - N}\(\phi_\pi + \nu
a_0\), \nnn \ell'^I &=& {\pp\ell^I\over\pp t^I} = {\pp\over\pp
  t^I}\ln\[|\eta(t^I)|^4(t^I + \t^I)\].\label{ake}\eea
Defining $\tilde B^m = B^m_c - B^m_Q$, the equations of motion for
$\omega_a$ and $a_0$ give
\bea \nabla^m B_m &=& - {1\over b_c}\l{\pp
     V\over\pp\omega_c}\r_{a_0,\omega_Q} - {1\over b'_Q}\l{\pp
     V\over\pp\omega_Q}\r_{a_0,\omega_c} = - {1\over b_c}\l{\pp
     V\over\pp\omega'_c}\r_{\omega'_Q} - {1\over b'_Q}\l{\pp
     V\over\pp\omega'_Q}\r_{\omega'_c},\nnn \nabla^m\tilde B_m &=& -
     {1\over b_c}\l{\pp V\over\pp\omega_c}\r_{a_0,\omega_Q} + {1\over
     b'_Q}\l{\pp V\over\pp\omega_Q}\r_{a_0,\omega_c} = - {1\over
     b_c}\l{\pp V\over\pp\omega'_c}\r_{\omega'_Q} + {1\over
     b'_Q}\l{\pp V\over\pp\omega'_Q}\r_{\omega'_c}, \nnn
     \Box a_0 &=& {b'_Q N\nu \over2(N_c - N)}\(\nabla^m B_m -
     \nabla^m\tilde B_m\) + \l{\pp V\over\pp
     a_0}\r_{\omega_c,\omega_Q} \eee - \nu\(\l{\pp
     V\over\pp\omega'_c}\r_{\omega'_Q} + {N_c\over N - N_c}\l{\pp
     V\over\pp\omega'_Q}\r_{\omega'_c}\),\label{oeom}\eea
where subscripted fields are held fixed.
The equations of motion for the three-form potentials $\Gamma = {}^*B$,
$\tilde\Gamma = {}^*\tilde B$, that are dual to the one-forms $B_m,\tilde
B_m$, give
\bea {k'\over\ell}B_m &=& - \nabla_m\(b_c\tilde\omega_c +
b'_Q\tilde\omega_Q\) - i\sum_I{b\over2}\(\ell'^I\pp_m t^I\mhc\), \nnn 0 &=& -
\nabla_m\(b_c\tilde\omega_c - b'_Q\tilde\omega_Q\)\quad\Rightarrow\quad
\tilde\omega_Q = {b_c\over b'_Q}\tilde\omega_c + \phi_0,\label{beom}
\eea
where $\phi_0$ is a constant phase. There are therefore two
independent neutral axions (besides $\im t^I$) that we can take to be
$a_0$ and 
\beq \omega = b_c\(\omega'_c + \nu a_0+ \phi_c\) = b_c\tilde\omega_c
= b'_Q\(\omega'_Q +{N_c\nu a_0\over N_c - N} + \phi_Q\)
\label{omega},\eeq
where $\phi_c$ and $\phi_Q$ are constant phases. Using 
\beq \l{\pp V\over\pp\omega}\r_{a_0} = {1\over b_c}\l{\pp
V\over\pp\omega'_c}\r_{\omega'_Q} + {1\over b_Q}\l{\pp
V\over\pp\omega'_Q}\r_{\omega'_c}\qquad \l{\pp V\over\pp a_0}\r_{\omega}
= - \nu\(\l{\pp V\over\pp\omega'_c}\r_{\omega'_Q} + {N_c\over N -
N_c}\l{\pp V\over\pp\omega'_Q}\r_{\omega'_c}\),\eeq
and combining \myref{oeom} and \myref{beom}, we obtain the equations
of motion of the dual scalar Lagrangian\footnote{It is straightforward
to check that all the equations of motion are the same as for the
original Lagrangian; see, for example~\cite{bgw}.}
\bea \L &=& - {2\ell\over k'}\(\pp_m\omega - \sum_I{b\over2}
\im\ell'^I\pp_m t^I\)\(\pp^m\omega - \sum_I{b\over2}\im\ell'^I\pp_m
t^I\) - V\mmm - \half\(\pp_m\bfs\pp^m\bfs + \pp_m\sigma_0
\pp^m\sigma_0 + \pp_m\bfa\pp^m\bfa + \pp_m a_0\pp^m a_0\) - K_{I\bar
J}\pp t^I\pp\t^{\bar J},\eea
with $V$ given by \myref{va} and \myref{omega}. There are two dynamical
degrees of freedom associated with the phases relevant for the strong
CP problem, the axion $\omega$ and the phase $\phi_\pi$.  Setting
the other fields at their vacuum values, \myref{va} takes the form
\beq V(\omega,\phi_\pi) = \tilde A_1 + \tilde A_2\cos\omega'_c +
\tilde A_3\cos\omega'_Q + \tilde A_4\cos(\omega'_c - \omega'_Q),
\label{vax}\eeq
where $\tilde A_1 = A_1(v) + m^2_0A_5(v,0),$ {\it etc.} The potential
\myref{vax} is minimized for
\beq 0 = {\pp V\over\pp\omega_a} = -\tilde A_2\sin\omega'_c - \tilde
A_4\sin(\omega'_c - \omega'_Q) = -\tilde A_3\sin\omega'_Q + \tilde
A_4\sin (\omega'_c - \omega'_Q).\eeq
This has CP conserving solutions $\omega'_a = 0,\pi$. There might
also be a CP violating solution $\omega'_a\ne0$ provided
\bea - 1 < {{\cos}\omega'_c} &=& -\half\({\tilde A_4\over \tilde A_3}
+ {\tilde A_3\over \tilde A_4} - {\tilde A_3 \tilde A_4\over \tilde
  A^2_2}\) < 1, \nnn - 1 < {{\cos}\omega'_Q} &=& -\half\({\tilde
  A_4\over \tilde A_2} + {\tilde A_2\over \tilde A_4} - {\tilde A_2
  \tilde A_4\over \tilde A^2_3}\) < 1, \eea
However the global minimum will occur for the CP conserving vacuum
with $\omega'_a = 0,\pi$ that maximizes the (negative) coefficients of
the two largest of $|\tilde A_{i\ne1}|$. For example if $\tilde
A_2>\tilde A_3>\tilde A_4>0$, the global minimum occurs\footnote{As
usual we fine-tune the dilaton K\"ahler potential to make the
potential positive semi-definite.} for $\myvev{V} = \tilde A_1 -
\tilde A_2 - \tilde A_3 + \tilde A_4.$ 

To study the effective theory for the axion and the mesons we can set
the moduli and the static condensates at their ground state values.  The
canonically normalized mesons $\sigma_0,\bfs,a_0,\bfa$ are defined as in
\myref{newpi4} and
\beq a = -\sqrt{2\ell\over k'}\omega\label{canax}\eeq
is the canonically normalized axion.\footnote{The sign is chosen
to give the standard coupling, \myref{alag} below; see (32) of
\cite{bg}.}   In the limit of vanishing meson masses, $m_0\to 0$,
the potential depends only on the scalars $\sigma$ and one linear
combination, mostly $a_0$, of neutral pseudoscalars:
\bea \sqrt{1 + c_a^2}\eta &=& {N_c-N\over\nu N}(\omega'_c - \omega'_Q
+ \phi_c - \phi_Q) = {a_0 + c_a a}\approx a_o, \nnn c_a &=&
{v\over\mu}{8\pi^2}\sqrt{k'\over N\ell}\(1 - {b'_Q\over b_c}\)
\ll1.\eea
The other (mostly axion) neutral pseudoscalar and the charged pseudoscalars
are massless Goldstone bosons of the nonanomalous
symmetry $SU(N)_L\otimes SU(N)_R\otimes U(1)$, where the nonanomalous
\uone\, is defined by \myref{kuone} and \myref{uone} with
\beq \alpha b_c = {1\over8\pi^2}\[\alpha\(N_c - N\) + 2\beta
N\]\label{nonan}.\eeq
If $m_0\ne0$, flavor-chiral \uone\, symmetry is broken, and there is no
longer the freedom to choose the R-parity of $Q$; in this case the
classical R-symmetry has $\beta = \alpha/2$, and it is anomalous at
the quantum level unless 
\beq b_c = {N_c\over8\pi^2} = {b'_Q N_c\over N_c - N}.\label{sym}\eeq
Writing 
\bea \omega'_c &=& - \nu\sqrt{1 + c_a^2}\eta - {\omega\over N
b_c}\(8\pi^2 b_c - N_c\) - \phi_c, \nnn \omega'_Q &=& - {\nu
N_c\sqrt{1 + c_a^2}\over N_c-N}\eta - {\omega\over N b_c}\(8\pi^2 b_c
- N_c\) - \phi_Q,\eea
we see explicitly that the potential \myref{va} depends only on one
neutral pseudoscalar $\eta$ at the point of enhanced symmetry
\myref{sym}.  With the \vev's $\myvev{\omega'_a}$ determined as
described above, the potential for the light pseudoscalars takes the form
\bea V(\bfa,a) &=& - c\Tr e^{i\mu(\bfa - c'_a
a/\sqrt{N})/v\sqrt{2}}\hc,\qquad c = m_0 \left|A_3(v) -
{A_2(v)\over|A_2(v)|}A_4(v)\r = {v^2\over\mu^2}m^2_{\bfa},\nnn c'_a
&=& {m_a\over m_{\bfa}} = \sqrt{k'\over N\ell}{v(8\pi^2 b_c -
N_c)\over\mu b_c}.\label{amass} \eea

For example, if we assume
$\rho_c\gg|u_c|,m_0|\bpi|$, and use the condition of (approximately)
vanishing vacuum energy~\cite{bgw}:
\beq k'/\ell\approx 3b_c^2/(1 + b_c\ell)^2,\label{vac}\eeq
the minimum of the potential when $m_0\to0$ is given by
\beq \myvev{\cos(\omega'_c -\omega'_Q)}={\alpha\over|\alpha|},\quad
\mu_v = v^{1\over2}\approx
\({\gamma\mu^2\Lambda^{3b_Q/b_1}\over\mG}\)^{1/2(2 + 3b_Q N/b_3}, \quad
\Lambda = \({b'_Q\lambda_Q\over4}\)^{b_1/3b_Q}\sim \Lambda_Q,\label{vev}\eeq
where 
\beq \alpha = {b_Q^\alpha\over b'_Q} - {3(b_c -
b'_Q)\over2b'_Q(1 + b_c\ell)}, \qquad \gamma = 
\sqrt{\alpha^2 + 3} - |\alpha|>0,\label{param}\eeq
and the $\eta,\sigma$ masses are
\beq m_{\eta} \approx 2\mG\sqrt{|\alpha|/\gamma} ,\qquad m_{\sigma_0}
\approx 2\mG\sqrt{6 - 2|\alpha|\gamma}/\gamma,\qquad m_{\bfs}
\approx 2\mG\sqrt{2 - |\alpha|\gamma}/\gamma,\label{msig}\eeq
where
\beq \mG\approx {b_c\over4}\rho_c\eeq
is the gravitino mass.  When $m_0$ is turned on there are small shifts
in the vacuum value \myref{vev} and the masses \myref{msig}, and
the pseudoscalars $\bfa,a$ acquire masses as in \myref{amass} with
\beq c =
4m_0^2{\Lambda^{7\over4}(\mG)^{3\over4}\mu^{1\over2}}\(3\myvev{\cos\omega'_Q}
- \beta\gamma \myvev{\cos\omega'_c}\),\qquad \beta = 1 - {3\over1+
b_c\ell},\label{vev2}\eeq
and the minimum is CP conserving with $\myvev{\cos\omega'_{Q,c}}=\pm1$
so as to make $c$ positive.  

The above supersymmetric model is not a realistic model for QCD for
several reasons.  The composite operators $u_Q$ and $\bpi\lowest$ are
composed of gauginos and squarks that get large masses proportional to
$\mG$, while the true light degrees of freedom are the quarks and
gauge bosons.  The corresponding composite operators are the axillary
fields $F_Q,\bff$ that have been eliminated by their equations of
motion.  We would like to trade the former for the latter.  More
precisely, the composite gauge fields are
\beq S\sim (F\cdot F)_{QCD}= - F_Q + u_Q\M\hc, \qquad P\sim
(F\cdot\tF)_{QCD} = 4\nabla^m B_m^Q.\eeq
The equation of motion for $F_a$ forces the coefficient of $S_a$ to
vanish, consistent with the definition $g^{-2}(\Lambda_a)=0$ of the
$\G_a$ scale $\Lambda_a$, and \myref{vyt} correctly
reproduces~\cite{bgw} the running of $g^{-2}$ from the string
scale to the condensation scale provided supersymmetry is unbroken
above that scale.  This is not the case for QCD, and the effective
``QCD'' Lagrangian \myref{lvyt} or \myref{lvyt3} is not valid below the scale
$\Lambda_c$ of supersymmetry breaking.  The arguments of the logs are
effective infra-red cut-offs. For gauginos and squarks, they should be
replaced by the actual masses, as was done in \myref{lvyt3} for quark
supermultiplets with masses above the QCD condensation scale.

The gaugino and squark mass terms
\beq \L_{\rm mass} = -
{k'\rho_c\over16\ell}\(e^{i\omega_c}\bar\lambda_R\lambda_L\hc\) - 
\mG^2|\tq|^2\eeq
are invariant under \myref{kuone} which is spontaneously broken by the
vacuum value $u_c\ne0$, but remains an exact
(nonlinearly realized) symmetry of the Lagrangian, since the
anomaly can be canceled by an axion shift as long as QCD
nonperturbative effects can be neglected.  Since $U_c$
transforms the same way as $U_Q$ an effective theory with the
correct anomaly structure under \myref{kuone} and \myref{uone}
is obtained by replacing \myref{lvyt3} by
\beq \L^Q_{VYT} = {1\over8}\superint{E\over R}U_Q\lbr
b_1\ln(e^{-K/2}U_Q) + b_2\ln(e^{-K/2}U_c) + b_3\[\ln(\det\bpi_n) -
\sum_{A=1}\ln M_A\]\rbr\hc,\label{lvyt4}\eeq
provided 
\beq b_3 = {1\over8\pi^2} = b^\alpha_Q,\qquad b_1 + b_2 = {N_c -
n\over8\pi^2} \equiv b'_n,\label{newb}\eeq
and we can choose $b_1$ and $b_2$ to better reflect the correct
infrared cut-offs for squarks and gauginos.  The potential is still of
the form \myref{va}, except that the functions $A_i$ depend on the
parameter $b_2/b_1$, which modifies the masses \myref{msig}, but the
axion mass is unchanged since it depends only on $b_1 + b_2 =
b'_n$. We may write the effective Lagrangian below the QCD scale in
terms of the quark condensate $\bff_n$ by using its equation of
motion. Putting everything except the light pseudoscalars at their
vacuum values, to leading order in $1/m_P$ we obtain
\bea \bfbf_n &\approx&  \(\sigma e^{-i\bfP'} - c_0\)e^{i\del}, \nnn
\bfP' &=& {\mu\over v\sqrt{2}}\(\bfa - {c'_a a\over\sqrt{n}}\) =
{\mu\over v\sqrt{2}}\bfa - {8\pi^2 b_c - N_c\over b_c
n}\sqrt{k'\over2\ell}a, \qquad c_0 = \mu^2m_0,\eea
and the effective potential for the light pseudoscalars takes the form
\beq c_0\sigma'e^{i\bfP'}\hc + O(m_0^2) = c\Tr\bff_n
\hc + O(m_0^2),\qquad c = c_0{\sigma'\over\sigma},\label{pot}\eeq
which is the standard result in QCD if $\Tr\bff$ is identified with
the quark condensate.  To check that this identification is correct,
we note that above the condensation scales the Lagrangian contains
the coupling
\beq \L\ni -
{a\over4}\sqrt{k'\over2\ell}\sum_a(F\cdot\tF)_a\equiv -
{a\over4F}\sum_a(F\cdot\tF)_a.\label{alag}\eeq
Under the K\"ahler \uone\, transformation \myref{kuone} and the
transformation \myref{uone} on the $n$ light quark supermultiplets,
the anomalies induce a shift
\beq \del\L\ni - {1\over4}\[\alpha b_c(F\cdot\tF)_c + (\alpha b'_n
+ 2n\beta_3)(F\cdot\tF)_Q\],\eeq
which is canceled in the nonanomalous case \myref{nonan} by the
axion shift
\beq a \to a - \alpha b_c\sqrt{2\ell\over k'},\eeq
This gives 
\beq \bff_n\to e^{i\alpha b_{\bff}}\bff_n, \qquad b_{\bff} ={ 8\pi^2 b_c -
N_c\over n},\eeq
which matches the phase transformation of the quark condensate:
\beq\chi_L\chi^c_L \to e^{i\alpha b_{\chi}}\chi_L\chi^c_L, \qquad b_\chi =
2{\beta\over\alpha} - 1 = {8\pi^2b_c - N_c + n\over n} - 1 =
b_{\bff}.\eeq
For $n=2$ we identify the factor $\exp(i\mu\bfa/v\sqrt{2})$
in the parametrization \myref{newpi4} with the operator $\Sigma =
e^{2i\pi^i T_i/F_\pi}$ of standard chiral Lagrangians, where $\pi^i$
are the canonically normalized pions, and $T_i$ is a generator of
$SU(2)$ normalized as in \myref{norm}. That is,
we identify $a_i$ with $\pi_i$ and
\beq \mu/v = 2/F_\pi,\qquad F_\pi \approx 93\MeV\label{fpi}\eeq 
giving
\beq m_a = {\left|8\pi^2 b_c - N_c\r\over b_c n}{\sqrt{n}F_\pi\over
F\sqrt{2}}m_\pi = {\sqrt{3}|8\pi^2b_c - N_c|F_\pi\over2\sqrt{n}(1 +
b_c\ell)}m_\pi,\label{ma}\eeq
where $F$ is the axion coupling defined in \myref{alag}, and we used
\myref{vac}.  If we assume $b_c\ell\ll1$, $b_c = .036$, which is the
preferred~\cite{bhn} value for LSP dark matter in the BGW
model~\cite{bgw}, we are very close to the symmetric point for $N_c =
3$: $8\pi^2b_c = 2.84$, so we get an (accidental) suppression of the
axion mass; if $b_c\ell\ll1$ and $n=2$:
\beq m_a\approx 5\times 10^{-13}\eV, \qquad F \approx 5.5\times
10^{19}\GeV.\label{bgn}\eeq
The value of $F$ is larger than the classical value with $k'
= 1/\ell,\;\ell = g^2/2\approx .25$, giving $F^{\rm class} \approx
1/2\sqrt{2} \approx 8.6\times 10^{17}\GeV,$ due to the corrections
to the dilaton K\"ahler potential needed for dilaton stabilization.
One also finds in the literature a different normalization for
the axion couping 
\beq \L\ni -{n a\over32\pi^2f_a}\sum_b(F\cdot\tF)_b,\qquad
f_a = {n F\over8\pi^2},\label{lit}\eeq
giving $f_a = 1.4\times 10^{18}\GeV$, and $f_a^{\rm
class} = 2.2\times 10^{16}$, in agreement with the calculation of~\cite{fpt}.
In models with an anomalous \uone, there are other factors that
determine the spectrum, and $b_c$ can be quite different.  In general
these factors tend to raise the scale of supersymmetry breaking
unless $b_c$ is smaller and/or $\ell$ is considerably larger than its
classical perturbative value $g^2_s/2\approx0.25$. Either of these 
would increase $f_a$ and probably increase $m_a$ by moving $b_c$ away
from the symmetric point.

The result \myref{ma} appears to differ from the standard result by a factor
$1- N_c/8\pi^2 b_c$.  However, $F$ is the axion coupling to Yang Mills
fields {\it above} the scale of supersymmetry breaking. We will see in the
next section that when we integrate out the gluinos of the supersymmetric
extension of the Standard Model, we generate a correction that modifies
the axion coupling strength $F^{-1}$ to $(F\tF)_Q$ by precisely that factor.

\mysection{QCD}\label{qcd}
In the real world squarks and gauginos are unconfined at the scale of
$\G_c$ condensation; they get masses proportional to $\mG$.
Therefore we integrate them out, as well as the dilaton and moduli, 
to get an effective theory for quarks and gauge bosons. For present
purposes, we can ignore the fact that quark masses come from Higgs
couplings and just take the quark superpotential and K\"ahler potential 
to be
\bea W_q(\hq,\hQ) &=& \eta^{-2}\(m e^{i\del}\hq^T \hq^c +
\hQ^TM \hQ^c\), \qquad K_q = \hq^{\dag}\hq + (\hq^c)^{\dag}\hq^c + 
\hQ^\dag\hQ + (\hQ^c)^\dag\hQ^c, \nnn
\hq^T &=& (\hq_1,\ldots,\hq_n),\qquad \hQ^T =
(\hQ_1,\ldots,\hQ_m),\qquad m + n = N.\eea
The chiral superfields $\hQ_A = (\tQ_A,Q_A,F_A)$ have masses
$M_A\gg\Lambda_{QCD}$, and we have used a nonanomalous $SU(N)$
transformation to make their mass matrix real and diagonal and to
diagonalize the mass matrix of the light quarks
$\hq_i = (\tq_i,q_i,f_i)$ which we take to have degenerate eigenvalues:
$|m_i| = m\ll\Lambda_{QCD}$. The relevant part of the Lagrangian at the
SUSY-breaking scale is (dropping kinetic terms for heavy fields and
setting the moduli at self-dual points)
\bea \L &=& \L_{\rm kin} - V - \L_Y,\qquad \chi = (q,q^c),\qquad \phi
   = (\tq,\tq^c),\qquad X = (Q,Q^c), \qquad \Phi = (\tQ,\tQ^c),\nnn
   \L_{\rm kin} &=& {k'\over4\ell}B^m B_m - b_c\omega_c\nabla^m B^c_m
   - {i\over2}\(\bar\chi_L\gamma^m D_m\chi_L\mhc\) -
   {1\over4g^2(\mG)}F\cdot F, \nnn V &=&
   {k'\over16\ell}\left|\rho_c e^{i\omega_c}(1 + b_c\ell) +
   \bar\lambda_R\lambda_L - 4\ell e^{K/2} W_q(\tq,\tQ)\right|^2 \mmm -
   {3\over16}\left|b_c\rho_c e^{i\omega_c} - 4 e^{K/2}W_q(\tq,\tQ)\r^2
   + F^T\bF + f^T\f, \nnn \bF_A &=& {1\over4}\[b_c\rho_c
   e^{i\omega_c}\Phi^\dag_A - 4 e^{K/2}W_q(\tq,\tQ)\Phi^\dag_A -
   4(M_0)_A\Phi_A\],\nnn \f_i &=& {1\over4}\[b_c\rho_c
   e^{i\omega_c}\phi^\dag_i - 4 e^{K/2}W_q(\tq,\tQ)\phi^\dag_i - 4 m_0
   e^{i\del}\phi_i\], \nnn \L_Y &=& Q^T M_0 Q^c + e^{i\del}m_0q^T q^c
   + \(\Phi^\dag X + \phi^\dag\chi\)^2W_q(\tq,\tQ)\mmm -
   i\sqrt{2}\[\chi^T(\lambda\cdot T)\bar\phi + X^T(\lambda\cdot
   T)\bar\Phi\]\hc, \nnn B_m &=& {}^*(d b + \Gamma)_m + \omega_m,
   \qquad D_m\chi = \D_m\chi + {i k'\over4}B_m\chi,
\label{lag}\eea
where $\D_m$ is a gauge covariant derivative, $M_0$ is defined
analogously to $m_0$, and $b_{m n}$,
$\Gamma_{m n p}$ and $\omega_m$ are, respectively, a two-form
potential, a three-from potential, and the Chern-Simons one-form for
unconfined gauge fields: $\nabla^m\omega_m = {1\over4}F\cdot\tF.$

If we neglect the small mass $m_0$, the Lagrangian \myref{lag} is
invariant under the nonanomalous transformation
\bea \lambda_a&\to& e^{i\alpha/2}\lambda_a,\quad \Phi_A\to
e^{i\alpha/2}\Phi_A, \quad X_A\to X_A, \quad \phi_i\to
e^{i\beta\alpha}\phi_i,\quad \chi_i\to e^{i\gamma\alpha}\chi_i,\nnn
\omega_c&\to& \omega_c + \alpha,\qquad \beta = {b_c -
b'_n\over2nb_3},\qquad \gamma = \beta - \half = {8\pi^2 b_c -
N_c\over2n},\label{fund2}\eea
where $b'_n$ and $b_3$ are defined as in \myref{newb}.  In
order to keep this approximate symmetry manifest in the low energy
effective theory, we redefine the squark and quark fields so as
to remove the $\omega_c$-dependence from all terms in the Lagrangian
for the heavy fields that do not involve the mass $m_0$:\footnote{We
are implicitly making invariant other heavy fields, such as the
dilatino ($\chi'_\ell = e^{i\omega_c}\chi_\ell$) and gravitino
($\psi'_\mu = e^{-i\omega_c}\psi_\mu$), that also transform under
\myref{fund2} in order to insure invariance of the full classical
Lagrangian.}
\beq \lambda_a= e^{i\omega_c/2}\lambda'_a,\quad \Phi_A=
e^{i\omega_c/2}\Phi'_A, \quad X_A= X'_A, \quad \phi_i=
e^{i\beta\omega_c}\phi'_i,\quad \chi_i= e^{i\gamma\omega_c}\chi'_i.
\label{redef}\eeq
The primed fields are invariant under \myref{fund2}, and when
expressed in terms of them, $V$ and $\L_Y$ have no dependence on
$\omega_c$ when $m_0\to0$; this assures that any effects of
integrating out the heavy fields will be suppressed by powers of
$m_0/M_A$, $m_0/\mG$ relative to the terms retained.  However,
these transformations induce new terms in the effective Lagrangian.
First, because the transformation \myref{redef} with $\omega_c$
held fixed is anomalous, it induces a term
\beq \L'\ni \Del\L = - {\omega_c b_c\over
4}(F\cdot\tF)_Q.\label{shift}\eeq
Secondly there are shifts in the kinetic terms; the ones that concern us here
are the fermion derivatives:
\beq \pp_m\lambda_L = e^{i\omega_c/2}\(\pp_m\lambda'_L +
{i\over2}\pp_m\omega_c\lambda'_L\),\qquad \pp_m\chi_L =
e^{i\gamma\omega_c}\(\pp_m\chi'_L + i\gamma\pp_m\omega_c\chi'_L\),\eeq
which corresponds to a shift in the axial connections $A_m$ in the fermion
connections: 
\beq \Del A_m^\lambda = - \half\pp_m\omega_c,\qquad \Del A_m^\chi = - 
\gamma\pp_m\omega_c.\label{dela}\eeq
Quantum corrections induce a nonlocal operator coupling
the axial connection to $F\tF$; at scales $\mu^2\sim\Box\gg
m^2_\lambda$ through the anomalous triangle diagram:
\beq \L_{\rm qu}\ni -
{1\over4}(F\cdot\tF)_Q{1\over\Box}\({N_c\over4\pi^2}\pp^m A_m^\lambda
+ {n\over2\pi^2}\pp^m A_m^\chi\).\label{tri}\eeq
The contribution to \myref{tri} from the shift \myref{dela} exactly
cancels the shift \myref{shift} in the tree level Lagrangian, leaving
the $\omega F\tF$ S-matrix element unchanged by the redefinition
\myref{redef}. However at scales $\mu^2\ll m^2_\lambda$, we replace
$\Box\to m^2_\lambda$ in the first term of \myref{tri} because the
contribution decouples, but the analogous contribution \myref{shift}
to the tree Lagrangian $\L'$ remains in the effective low energy
Lagrangian.  This is a reflection of the fact that the classical
symmetry \myref{fund2} of the unprimed variables is anomalous.  The
gluino contribution to that anomaly is not canceled by the gluino
mass term, because the gluino mass does not break the symmetry; its
phase $\omega_c$ is undetermined above $\Lambda_{QCD}$ and transforms
so as to make the mass term invariant.  To see that
the gluino contribution to the anomaly does not decouple, we write the
(unprimed) gaugino contribution to the one-loop action as
\beq S_1 = - {i\over2}\Tr\ln(i\notD + m_\lambda) = S_A +
S_N,\label{s1}\eeq
where
\beq S_A = - {i\over2}\Tr\ln(i\notD)\eeq
is mass-independent and contains the gaugino contribution to the
anomaly:
\beq \del\L \ni \del S_A = - {\alpha
N_c\over32\pi^2}(F\cdot\tF)_Q.\label{dell}\eeq
The mass-dependent piece
\beq S_N = - {i\over2}\Tr\ln(-i\notD + m_\lambda) + 
{i\over2}\Tr\ln(-i\notD)\eeq
is finite and therefore nonanomalous. A constant mass term would break
the symmetry and the contribution from $S_N$ would exactly cancel that
from $S_A$ in the limit $\mu/m_\lambda\to0$. However it clear that
$S_N$ is invariant under \myref{kuone} because the gaugino mass is
covariant. In \myapp{anomap} we explicitly show by direct calculation
that gaugino loops give the contribution \myref{dell} under
\myref{fund2} in the limit $m_\lambda\gg\mu$, which in this limit
arises only from the phase of the mass matrix.  This implies that the
effective low energy theory must contain a coupling
\beq \L_{\rm eff}\ni\L_{\rm anom}= - {\omega_c
N_c\over32\pi^2}(F\cdot\tF)_Q,\label{lanom}\eeq
which is precisely the term that is generated by the redefinitions in
\myref{redef}.  The difference $(S_1)_\lambda - (S_1)_{\lambda'}$ in
the one-loop actions calculated with primed and unprimed gaugino
fields is just given by the first expression for $\del S_1$ in
\myref{dels1}, in the limit of small $\omega\to\alpha$, which can
trivially be integrated to include arbitrary $\omega$.  In order to
respect the full classical invariance of the Lagrangian, we have to
include the transformation on the squarks and quarks, giving the
effective coupling in \myref{shift}.

Setting the squarks and gauginos, as well as the heavy quarks, to zero
gives the effective light field tree Lagrangian at a scale
$\Lambda_{QCD}<\mu< M_A$:
\bea \L &=& \L_{\rm kin} - V - \L_Y + \Del\L,\qquad V
   = {1\over16\ell}\[k'(1 + b_c\ell)^2 - 3b_c^2\ell\]\rho^2_c, \nnn
   \L_{\rm kin} &=& {k'\over4\ell}B^m B_m - b_c\omega_c\nabla^m B^c_m
   - {i\over2}\(\bar\chi'\gamma^m D_m\chi'\mhc\)-
   \sum_a{1\over4g_a^2(\mu)}(F\cdot F)_a, \nnn D_m\chi'_L &=&
   \D_m\chi'_L + {i k'\over4}B_m\chi'_L +
   i\gamma\pp_m\omega_c\chi'_L,\qquad \L_Y = e^{i(\del +
   2\gamma\omega_c)}m_0\bar q'_R q'^c_L\hc,\nnn \nabla^m B_m &=&
   \nabla^m B^c_m + {1\over4}F\cdot\tF = \nabla^m({}^*\Gamma)_m +
   {1\over4}F\cdot\tF.
\label{lag2}\eea
The equations of motion for $\Gamma$ and $\omega_c$ give (the equation for
$b_{mn}$ is redundant; this field can be absorbed into $\Gamma$ by
a gauge transformation $\Gamma\to\Gamma - d b$)
\bea 0 &=& {k'\over2\ell}B_m + b_c\nabla_m\omega_c + {k'\over2}j_m
\qquad j_m = \bar\chi'_L\sigma_m\gamma_5\chi'_L,\nnn 0
&=& b_c\nabla^m B^c_m + {\pp\over\pp\omega_c}\L_Y - 2\gamma\nabla^m j_m +
{b_c\over4}(F\cdot\tF)_Q\eee b_c\(\nabla^m B_m - {1\over4}F\cdot\tF\)
+ {\pp\over\pp\omega_c}\L_Y - 2\gamma\nabla^m j_m + {b_c\over4}(F\cdot\tF)_Q\eee -
b_c\[\nabla^m\({2\ell\over k'}b_c\nabla_m\omega_c + \ell j_m\) +
{1\over4}F\cdot\tF\]\mmm + {\pp\over\pp\omega_c}\L_Y  +
2\gamma\nabla^m j_m + {b_c\over4}(F\cdot\tF)_Q, \eea
which is the scalar equation of motion for the equivalent Lagrangian
\bea \L &=& - {\ell\over k'}\pp^m\omega\pp_m\omega -
   {i\over2}\(\bar\chi'_L\gamma^m D_m\chi'_L\mhc\) - \[e^{i(\del +
   2\gamma\omega/b_c)}m_0q'^T q'^c\hc\]\mmm -
   \sum_a{1\over4g_a^2(\mu)}(F\cdot F)_a + {\omega\over4}[F\cdot\tF -
(F\cdot\tF)_Q], \nnn D_m\chi'_L &=& \D_m\chi'_L +
   {i\over2b_c}\(2\gamma - b_c\ell\)\pp_m\omega\chi'_L, \qquad \omega =
   b_c\omega_c.\label{lag3}\eea
If we ignore $m_0$ the QCD part of \myref{lag3} is invariant under a
shift in $\omega$ by a constant, which is the same as the nonanomalous
symmetry \myref{fund2}, after the redefinitions \myref{redef}.

From now on we drop the primes on the quark fields.  We define the
canonically normalized axion as in \myref{canax} and include explicitly
the QCD instanton-induced term, since it cannot be treated as a total
derivative when we approach the QCD scale where we obtain an effective
Lagrangian for pseudoscalars. Then, in terms of four-component Dirac
spinors, the Lagrangian \myref{lag3} reads,
\bea \L &=& - i\bar q\notD q - m_0\(e^{i(\del - a/f)}\bar q_R q_L\hc\)
- \half\pp_m a\pp^m a\mmm - {1\over4}\(g_\gamma^{-2}(\mu)F^2 +
\sqrt{k'\over2\ell}a F\cdot\tF\)_\gamma - {1\over4}\(g_Q^{-2}(\mu)F^2
+ {\theta\over8\pi^2}F\cdot\tF\)_Q,\nnn \notD q &=& \notcD q -
{i\over2}\sqrt{k'\ell\over2}\(1 - {2\gamma\over b_c\ell}\)\notd
a\gamma_5q,\qquad f^{-1} = \sqrt{k'\over2\ell}
{8\pi^2b_c - N_c\over n b_c},\label{lqg}
\eea
where the subscript $\gamma$ stands for\footnote{We have ignored 
a shift in the axion coupling to $(F\tF)_\gamma$ analogous to
\myref{shift} which is canceled in the S-matrix element
by the loop contribution analogous to \myref{tri}.} QED.  The first
term in the quark connection is the standard one; in the classical
limit $k' = \ell^{-1}$ it reduces to $- i\gamma_5 a/2\sqrt{2} =
-i\gamma_5\im s/4\re s$.  The second term is a result of the quark
field redefinition in \myref{redef}. The Lagrangian \myref{lqg} has a
classical symmetry
\beq a\to a + \alpha,\qquad q\to e^{-i\gamma_5\alpha/f}q\eeq
that is anomalous unless (neglecting $\del\L_{QED}$) $f^{-1} =0$,
which is just the condition \myref{sym} found previously.
We can now make an anomalous chiral transformation on the
quarks to remove the $\theta$ term from \myref{lqg}:\footnote{See
for example the discussion in Ch. 23.6 of~\cite{wein}.}
\beq q \to e^{i\theta\gamma_5/n}q.\eeq
Below the QCD confinement scale the physical degrees of freedom
are the pions; with the usual parameterization
\beq v^3e^{i\phi}\Sigma^a_b = v^3 (e^{i(\phi+2\bfpi/F_\pi)})^a_b \sim
q^a_L(\bar q_R)_b, \qquad \bfpi = \half\sum_i\pi^i\lambda_i,
\qquad \bigvev{\bfpi}=0 .\eeq
Using standard chiral symmetry arguments
we get the effective Lagrangian
\bea \L_{eff} &=& {1\over4}F_\pi^2\Tr\(\pp_\mu\Sigma\pp^\mu\Sigma^{\dag}\) +
\half\pp_\mu a\pp^\mu a + i F_\pi^2\sqrt{k'\ell\over2}\(1 -
{2\gamma/b_c\ell}\)\pp_\mu a\Tr\{\Sigma,\pp^\mu\Sigma^{\dag}\}\mmm +
\half\lambda v^3\[\Tr(e^{-i a'/f}\Sigma
m_0)\hc\],\qquad a' = a - f(\del + \phi - 2\theta/n).\label{leff}\eea
The third term on the RHS's of \myref{leff} is the coupling of the
universal axion to the chiral \uone\, Noether current as implied by
\myref{lqg}. Since Tr$\bfpi= 0$ it does not introduce any mixing of
the axions with the pions. The potential for the light pseudoscalars
$a,\bfpi$ is identical to that in \myref{pot}.  Since since
$m_0\myvev{\Sigma}$ is real the potential is an even function of $a'$
and has a minimum at $\myvev{a'} = 0$, so CP is conserved. If we take
$n=2$ and allow for $m_u\ne m_d$, the mass terms are
\bea V_m &=& \half\lambda v^2\[\(m_u + m_d\)\({\vec\pi^2\over F_\pi^2} +
{a'^2\over f^2}\) - 2\(m_u - m_d\){a'\pi^0\over f F_\pi}\]\nnn &\approx&
\half m^2_\pi\[2\pi^+\pi^- + \(\pi^0 + {F_\pi(1 -z)\over
f(1+z)}a'\)^2\] + \half m^2_a a_0^2,\eea
where
\beq a_0 \approx a' - {F_\pi(1 -z)\over f(1+z)}\pi^0, \qquad m_a
\approx 2m_\pi{F_\pi\sqrt{z}\over f(1+z)}, \qquad z = {m_u\over
m_d}.\eeq
To see that this\footnote{The coupling constant $f_a$ used
in~\cite{sred} is a factor two larger than the one defined in
\myref{lit} and used in~\cite{fpt}.  Taking this into account we agree
with~\cite{sred}, but differ by a factor two with~\cite{fpt}
and~\cite{wein}. The latter uses $F_\pi = 186\MeV$, and factors of two
are missing in the arguments of substitutions $\bar q q\to
f(\pi^0/F_\pi)$.} is the standard result~\cite{sred} for $n=2$, we
note that if we undo the redefinition of the quarks in \myref{redef}
by a transformation $q\to q'' = e^{-ia\gamma_5/2f}q$, we put back a
term
\beq \L\ni - {n a\over32\pi^2f}(F\cdot\tF)_Q = - \sqrt{k'\over2\ell}
\(1 - {b_0\over b_c}\){a\over4}(F\cdot\tF)_Q,\label{newq}\eeq
which differs from the universal axion coupling \myref{alag} by the
gaugino contribution \myref{lanom} that is generated when the
gauginos are integrated out.

\mysection{CP violation}\label{cpv} 
At the point of enhanced symmetry $b_c = 8\pi^2N_c$, the nonanomalous
symmetry \myref{fund2} does not include a chiral transformation on the
quarks, and one loses the solution to the CP problem.  The axion
decouples from the quarks in the effective Lagrangian \myref{lqg}, and
its \vev\, cannot be adjusted to make the quark mass matrix real in
the $\theta = 0$ basis.  There is no reason to expect that nature sits
at this point, but if the axion mass is very small one should worry
about other sources of an axion potential, such as higher dimension
operators~\cite{bd1}.  These were studied in~\cite{bg} in the context
of the modular invariant gaugino condensation models considered here.
Modular invariance severely restricts the allowed couplings; the leading
contribution to the axion mass takes the form
\beq m'^2_a \approx
{p^3|u|^2k'\lambda|\eta^2e^{-K/2}u|^p\over4b_c^2\ell}\[3b_c - (1 + b_c\ell)
{k'}\]\label{hdmass}\eeq
where $\lambda$ is a dimensionless coupling constant, and $p$ is the
smallest integer allowed by T-duality.  An orbifold compactification
model with three complex moduli and an $[SL(2,{\bf Z})]^3$ symmetry has
$p=12$, and, with the values of the various parameters used above and
$\lambda\approx1$, one finds $m'_a \approx 10^{-63}$eV, which is
completely negligible.  However if the symmetry is restricted, for
example, to just $SL(2,{\bf Z})$ one has $p=4$ and the contribution
from \myref{hdmass} is of the order of \myref{bgn}.  The axion potential 
is now
\beq V(a) = - f^2m^2_a\cos(a/f + \phi_0) - f'^2m'^2_a\cos(a/f'),\qquad
f'^{-1} = {p\over b_c}F^{-1} = {p n\over8\pi^2b_c
-N_c}f^{-1},\label{cp}\eeq
where we have absorbed constant phases in $a$ and/or in $\phi_0 =
-(\del + \phi + 2\theta/n)$ so as to make the coefficients negative.
The strong CP problem is avoided if for some value of $b_c$ the vacuum
has $\myvev{\bar\theta} = \myvev{n(a/f + \phi_0)/2}<10^{-9}$ for any
value of $\phi_0$.  For values of $b_c$ in the
preferred~\cite{gn2,bhn} range $.3\le b_c \le .4$ this does not occur.
For example for $b_c = .036$ with $p=4$ and $f'/f\approx1/50$, this
requires $f'^2m'^2_a/f^2m^2_a<10^{-10}$, whereas evaluating \myref{cp}
gives $f'^2m'^2_a/f^2m^2_a\approx 4\times10^{-4}$ in this case.  A
numerical analysis shows that the CP problem is avoided provided $p\ge 5$,
that is, provided the T-duality group is
not the minimal one, which is the case for most compactifications of the
weakly coupled heterotic string.

\mysection{Conclusions}\label{conc}
We have shown that there is an enhanced symmetry point where the 
universal string axion
mass vanishes even in the presence of quark masses in 
string-derived models where
supersymmetry is broken by condensation of a simple gauge group in a
hidden sector.  As a consequence, the axion mass can be suppressed
relative to conventional estimates~\cite{fpt} for the mass of the
string axion.  The conditions under which
the universal axion can serve as the Peccei-Quinn axion were examined
and it was found that the strong CP problem is avoided for all but
the minimal $SL(2,{\bf Z})$ version of the T-duality group of the
weakly coupled heterotic string. Most compactifications have a larger
T-duality group, and from a phenomenological point of view, a larger
group is desirable for generating~\cite{rp} the R-parity of the MSSM.
Although our results were obtained using the linear
supermultiplet formulation for the dilaton superfield, we expect
that they can be reproduced in the chiral multiplet formulation.
The implications of our results for cosmological observation will be
presented elsewhere. 

\vskip 0.20in
\noindent {\bf \Large Acknowledgments}

\vspace{5pt}

\noindent 
       We wish to thank Joel Giedt, Dan Butter and Soojin Kwon for
helpful correspondence and discussions.  This work was supported in
part by the Director, Office of Science, Office of High Energy and
Nuclear Physics, Division of High Energy Physics of the
U.S. Department of Energy under Contract DE-AC02-05CH11231, in part by
the National Science Foundation under grant PHY-0098840.

\myappendix

\mysection{Anomaly}\label{anomap}
Consider the tree Lagrangian for a Majorana fermion $\lambda$
\bea \L &=& -{i\over2}\bl\notD\lambda - {\half}\bl_L\m\lambda_R\hc,
\qquad D_m = \D_m + iA_m\gamma_5, \qquad \D_m = \pp_m + iT\cdot
a_m , \nnn A_m &=& \half\pp_m\omega, \qquad \m = m^\dag =
e^{i\omega}\mu,\label{lmaj}\eea
where $a_m$ is a gauge field.  The classical Lagrangian is invariant
under
\beq\lambda_L\to e^{i\alpha/2}\lambda_L,\qquad \omega\to\omega +
\alpha.\label{class}\eeq
This symmetry is broken at the quantum level by the anomaly:
\beq \del\L\ni - {\alpha\over32\pi^2}F\cdot\tF.\label{anom}\eeq
For $\mu\to0$ this is just determined by the standard triangle diagram
for the $a^2A$ three-point function.  If the mass of $\lambda$ were constant
it would explicitly break the symmetry \myref{class} and the explicit
breaking would exactly cancel the anomalous breaking, given no contribution
 to the 3-point function at momentum scales $|p^2|\ll\mu^2$.  However
since the mass term in \myref{lmaj} respects the symmetry, the
contribution to \myref{anom} is independent of the mass parameter $\mu$.
Here we show this explicitly in the limit of very large mass, using
the methods of~\cite{gjs}.  Under \myref{class} the effective action
\myref{s1} changes by
\bea \del S_1 &=& - {i\over2}\Tr\ln(i\notD - \gamma_5\del\notA +
m_\lambda + \del m) + {i\over2}\Tr\ln(i\notD + m_\lambda) \eee -
{i\over2}\int{d^4p\over(2\pi)^4}\sum_{n=0}^\infty\Tr(-{\cal
R})^n\del{\cal R},\nnn {\cal R} &=& -{1\over p^2 - \mu^2}\(\{p^m,G_m\} +
G^m G_m - {i\over2} \sigma\cdot\hG +
i\widehat{[\notD,m_\lambda]}\),\qquad\sigma_{m n} =
{i\over2}[\gamma_m,\gamma_n],\nnn \del{\cal R} &=& - {1\over p^2 -
\mu^2}\(\notp - m_\lambda\)\(\del\notA - \del m_\lambda\), \nnn m_\lambda &=&
\mu e^{i\omega\gamma_5}, \qquad \hat f = e^{-iD\cdot\pp/\pp p}
f(x)e^{iD\cdot\pp/\pp p}, \qquad G_{m n} = [D_m,D_n],\label{dels1}\eea
and $G_m$ is also an expansion in the operator ${D\cdot\pp/\pp p}$
acting on $G_{m n}$.  For $\mu\to0$, each term in the sum is infrared
divergent, and the derivative expansion must be resummed to give
\myref{anom}. For large $\mu$ the only contribution to \myref{anom}
that is not proportional to an inverse power of $\mu^2$ involves
$\del m_\lambda = i\gamma_5\alpha m_\lambda$ and $G_{m n}\ni iF_{m n}$:
\bea \del S_1&\ni& - {i\over2}\int{d^4p\over(2\pi)^4}\Tr{\cal
 R}^2\del{\cal R} \ni {i\over8}\int{d^4p\over(2\pi)^4}\Tr{(\sigma\cdot
 G)^2m_\lambda\del m_\lambda \over (p^2 + \mu^2)^3}\eee -
 \int{ip^2dp^2\over16\pi^2}{\mu^2i\alpha\over (p^2 +
 \mu^2)^3}G\cdot\tG = - {\alpha\over32\pi^2}F\cdot\tF.\eea
In the models considered the symmetry is only global, $\alpha =$
constant in \myref{class}, but this has no bearing on the calculation
or the anomaly.  When we sum over all the gaugino contributions we get a
factor $N_c$.

\end{document}